\documentclass[aps,prl,twocolumn,superscriptaddress,floatfix,10pt]{revtex4-1}
\usepackage{amssymb,amsmath}
\usepackage{graphicx}
\usepackage{float}
\usepackage{bm}
\usepackage{multirow}
\usepackage{dcolumn}
\usepackage{tabularx}
\usepackage{longtable}
\usepackage[dvipsnames,usenames]{color}
\usepackage[latin1]{inputenc}

\newlength{\dbarheight}

\begin{document}

\title{Switchable and tunable Rashba-type spin splitting in covalent perovskite oxides}

\author{Julien Varignon} \affiliation{Unit\'e Mixte de Physique, CNRS,
  Thales, Universit\'e Paris Sud, Universit\'e Paris-Saclay, 91767,
  Palaiseau, France} \email{julien.varignon@cnrs-thales.fr}

\author{Jacobo Santamaria} \affiliation{Laboratorio de
  Heteroestructuras con aplicacion en Spintronica, Unidad Asociada
  CSIC/Universidad Complutense de Madrid, Sor Juana In\'es de la Cruz,
  3, 28049 Madrid, Spain, Instituto de Magnetismo Aplicado,
  Universidad Complutense de Madrid, 28040 Madrid
  Spain}\affiliation{Unit\'e Mixte de Physique, CNRS, Thales,
  Universit\'e Paris Sud, Universit\'e Paris-Saclay, 91767, Palaiseau,
  France}

\author{Manuel Bibes} \affiliation{Unit\'e Mixte de Physique, CNRS,
  Thales, Universit\'e Paris Sud, Universit\'e Paris-Saclay, 91767,
  Palaiseau, France}

\date{\today}
%========================================================================= 
%========================================================================= 
%=========================================================================
%========================================================================= 
\begin{abstract}
In transition metal perovskites (ABO$_3$) most physical properties are
tunable by structural parameters such as the rotation of the BO$_6$
octahedra. Examples include the N\'eel temperature of orthoferrites, the
conductivity of mixed-valence manganites, or the band gap of
rare-earth scandates. Since oxides often host large internal electric
dipoles and can accommodate heavy elements, they also emerge as prime
candidates to display Rashba spin-orbit coupling, through which charge
and spin currents may be efficiently interconverted. However, despite
a few experimental reports in SrTiO$_3$-based interface systems, the
Rashba interaction has been little studied in these materials, and its
interplay with structural distortions remain unknown. In this Letter,
we identify a bismuth-based perovskite with a giant,
electrically-switchable Rashba interaction whose amplitude can be
controlled by both the ferroelectric polarization and the breathing
mode of oxygen octahedra. This particular structural parameter arises
from the strongly covalent nature of the Bi-O bonds, reminiscent of
the situation in perovskite nickelates. Our results not only provide
novel strategies to craft agile spin-charge converters but also
highlight the relevance of covalence as a powerful handle to design
emerging properties in complex oxides.
\end{abstract}
\maketitle

Spintronics exploits the spin degree of freedom of carriers in
addition to their charge, giving rise to a very broad range of
electronic applications including magnetic memories or magnetic
sensors~\cite{RevueChappert,Arxiv1512.05428}. It relies on the
exchange interaction of electrons with local magnetic moments of a
ferromagnet to generate spin currents. Usually, the ferromagnets are
simple transition metal elements such as Co, Ni or their
alloys. Nevertheless, the control of their magnetization typically
requires large energies and they generate undesired magnetic fields
hindering high densification of devices. In the search for lower power
spintronic devices, the spin-orbit interaction (SOI) has been
identified as a promising pathway to achieve very efficient
spin-to-charge current conversion and {\em vice versa}. The two
phenomena at the core of this branch of spintronics, also called
spin-orbitronics, are the spin Hall effect
(SHE)~\cite{SHE-prediction,SHE-semiconducteur1} and inverse spin Hall
effect (ISHE)~\cite{ISHE-observation,ISHE-Si}. For instance, the
former effect allows efficient charge-to-spin current conversion in
non-magnetic materials based on heavy elements~\cite{Liu-Science336},
therefore alleviating the need for ferromagnets to generate spin
currents.

More recently, renewed interest for SOI based applications arise
thanks to the Rashba effect~\cite{Rashba}. In materials lacking
inversion symmetry, the interplay between the electric field and the
SOI lifts the degeneracy of electronic bands according to their
spin. For a material exhibiting a polar axis oriented along the $z$
direction, Rashba proposed the following spin-orbit interaction:
\begin{equation}
  H_R = \alpha_R (\sigma_x k_y - \sigma_y k_x)
  \label{eq1}
\end{equation}
where $\alpha_R$ is the Rashba coefficient -- proportional to the
spin-orbit strength of the material $\lambda_{SO}$ and to the
magnitude of the electric field --, $\sigma_i$ (i=x,y) are Pauli
matrices and $k_j$ (j=x,y) is the momentum of the electron. The Rashba
interaction leads to a spin-locking of electrons according to their
momentum $\vec k$ enabling efficient spin-to-charge current
interconversions through Edelstein and inverse Edelstein
effects~\cite{EffetEdelstein,EffetEdelsteinInverse}. Although
originally formulated at interfaces and
surfaces~\cite{BiAg,Revue-Ge111,GePb,LAO-STO-Edouard}, the Rashba
interaction is now generalized to non-centrosymmetric bulk
materials~\cite{BiTeI,GeTe,BiAlO3}, although it may coexist with the
Dresselhaus interaction~\cite{Dresselhaus}.

In the context of SOI related devices, ABO$_3$ oxide perovskites
remain largely unexplored. They encompass a wealth of properties --
including ferroelectricity and magnetism -- and functionalities
originating from the coupling of their structural, electronic and
magnetic degrees of freedom~\cite{Zubko2011interface}. Most notably,
the lattice distortions usually control the properties of the
perovskite, such as the metal-insulator phase transition temperature
in rare-earth nickelates~\cite{RevueNickelates1,NickelatePhilippe},
the Néel temperature of
orthoferrites~\cite{Orthoferrites1,Orthoferrites2} or the band gap
value of rare-earth scandates~\cite{Scandates1,Scandates2}. Being
largely ionic in most cases, many oxide perovskites host large
internal electric dipoles. Moreover, they can also accommodate heavy
elements, an important aspect for SOI based applications. So far,
BiAlO$_3$ has been proposed theoretically to be the first oxide
exhibiting ferroelectricity and Rashba physics~\cite{BiAlO3} --
although this coexistence was already observed in the semiconductor
GeTe~\cite{GeTe}.  While BiAlO$_3$ harbors a heavy element and a large
ferroelectric polarization -- estimated around 90~$\mu$C.cm$^{-2}$ --,
the computed Rashba coefficient $\alpha_R$ remains relatively moderate
-- around 0.39 and 0.74 eV.\AA -- although comparable to conventional
systems such as (111) oriented Bi~\cite{Bi111}.
In fact, other ingredients such as a narrow band gap and/or similar
atomic and orbital characters for the electronic states located around
the Fermi level are proposed to be additional key ingredients for
reaching large $\alpha_R$~\cite{BiTeI-PRB84}. Covalent oxides that are
characterized by a strong hybridization of B cations and O anions
electronic states around the Fermi level, are therefore an ideal
platform to engineer materials undergoing large Rashba
effects. Furthermore, oxides are usually characterized by numerous
lattice distortions but the role of the latter on the Rashba physics
remains unexplored.

In this letter, we use {\em first-principles} simulations to identify
the existence of large Rashba effects with $\alpha_R$ as large as 8
eV.\AA~in a ferroelectric phase of the covalent oxide strontium
bismuthate SrBiO$_3$ reachable by epitaxial strain. We further reveal
that the amplitude of the Rashba coefficient is tunable by a lattice
distortion strongly connected to bands dispersion and to the level of
hybridization between B and O electronic states. Our results therefore
highlight a new pathway to control and optimize the Rashba interaction
in oxides, further revealing the potential of these materials for
spin-orbitronics.

\begin{figure}
\begin{center}
\resizebox{8.0cm}{!}{\includegraphics{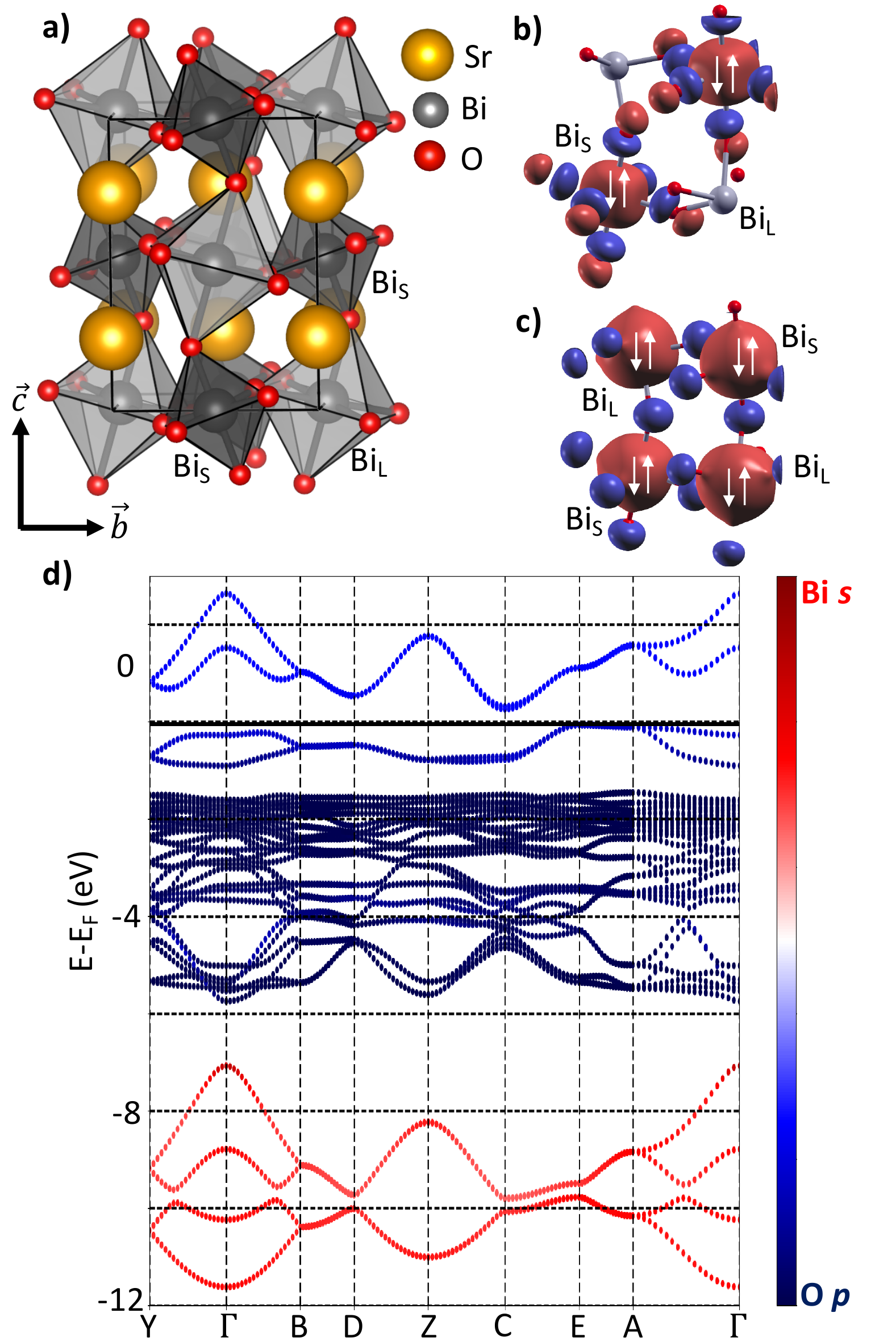}}
\end{center}
\caption{Structural and electronic properties of SrBiO$_3$ ground
  state. a) Ground state $P2_1/n$ monoclinic structure. b) $s$-like
  Wannier functions associated with the 2 bands lying above the Fermi
  level. c) $s$-like Wannier functions associated with the 4 low lying
  bands between -12 eV and -6 eV. d) Angular momentum resolved band
  structure of the monoclinic $P2_1/n$ phase. The Fermi level $E_F$ is
  set to 0. The high symmetry coordinates in the Brillouin zone of the
  $P2_1/n$ symmetry are the following: $\Gamma$ (0,0,0), B (1/2,0,0),
  Y (0, 1/2,0), A (1/2,1/2,0), Z (0,0,1/2), C (0,1/2,1/2),
  E(1/2,1/2,1/2) and D(1/2,0,1/2).}
\label{fig1}
\end{figure}

Strontium bismuthate Sr$^{2+}$Bi$^{4+}$O$_3$ is an insulator adopting
a monoclinic $P2_1/n$ symmetry at low
temperature~\cite{HowardActaB,SBO-Bordet}. The ground state structure
is characterized by the usual oxygen cage rotations appearing in
perovskites, $a^-a^-c^+$ antiferrodistortive motions in Glazer's
notation, as well as by a breathing of the oxygen cage octahedra
distortion B$_{\text{oc}}$ (see
Figure~\ref{fig1}.a)~\cite{HowardActaB,SBO-Bordet}. The latter
distortion produces a rocksalt-like pattern of large and small O$_6$
groups resulting in Bi sites splitting. In the following, Bi cations
sitting in compressed (extended) O$_6$ groups are labeled Bi$_{\rm S}$
(Bi$_{\rm L}$) (see Figure~\ref{fig1}.a). Since SrBiO$_3$ shares
several features with rare-earth nickelates RNiO$_3$, the appearance
of B$_{\text{oc}}$ in bismuthates is sometimes interpreted in terms of
ionic charge disproportionation of unstable Bi$^{4+}$ valence state to
the more stable Bi$_{\rm L}^{3+}$ + Bi$_{\rm S}^{3+}$ electronic
configurations in the insulating
phase~\cite{Disp1,BBO-IR-Lobo1,BBO-IR-Lobo2}. Alternatively, it is
also proposed to be a consequence of charge transfer effects from O
anions to Bi cations~\cite{BBO-Sawatzky}. Nevertheless, Tonnhauser and
Rabe demonstrated that the appearance of B$_{\text{oc}}$ is
responsible of the band gap opening in bismuthates \cite{Rabe-BBO}.

Although very similar to rare-earth nickelates, bismuthates also
exhibit many other striking properties such as high T$_{\rm c}$
superconductivity upon hole
doping~\cite{BBO-HighTc-K,BBO-HighTc-Pb,BBO-HighTc-Pb-Optical,SBO-HighTc,SBO-Bordet}
or possible topological surface states accessible upon electron
doping~\cite{NatPhysicsBBO}. The latter point highlights the strength
of the SOI in these materials, being {\em de facto} an ideal
playground to study theoretically the interplay between lattice
distortions, covalence and spin-orbit related phenomena.

{\em First-principles} calculations were carried out with Density
Functional Theory (DFT) using the Vienna {\em Ab initio} Simulation
Package (VASP)~\cite{VASP1,VASP2}. We used the PBE functional revised
for solids~\cite{PBEsol} with a cutoff energy of 500 eV and a $6\times
6 \times 4$ ($4\times 4 \times 4$) Monkhorst-Pack kpoint mesh for 20
(40) atoms unit cells. We used Projector Augmented Waves (PAW)
pseudopotentials~\cite{PAW} for core electrons with the following
valence electron configurations: 4s$^2$4p$^6$5s$^2$ (Sr), 6s$^2$6p$^3$
(Bi) and 2s$^2$2p$^4$ (O).  Geometry optimizations were performed
until forces are lower than 0.01 meV.\AA$^{-1}$ without spin-orbit
interaction included unless stated. Born effective charges and phonons
frequencies were computed with the density functional perturbation
theory~\cite{RMP-Baroni,Gonze-97}. Band structures and spin textures
were plotted with the PyProcar script~\cite{PyProcar}. Maxi-Localized
Wannier Functions (MLWFs) were extracted with the Wannier90
package~\cite{Wannier90-1,Wannier90-2,Wannier90-3}.

We first inspect the properties of bulk strontium bismuthate.
Geometry relaxation yields an insulating monoclinic $P2_1/n$ ground
state with a band gap of 0.3 eV, other polymorphs observed
experimentally at higher temperature -- $Fm\bar{3}m$, $R\bar{3}c$ or
$I_2/m$ symmetries -- are found higher in energies. Our optimized
lattice parameters (a=5.9400~\AA, b=6.1365~\AA, c=8.4859~\AA~and
$\beta$=90.0597$^{\circ}$) are found in very good agreement with
experiments from reference~\onlinecite{SBO-Bordet}, yielding 0.55 \%
of error on the total volume. Regarding the electronic structure of
the material, we extract for both Bi$_{\rm S}$ and Bi$_{\rm L}$
cations similar site occupancies N, although a weak charge order does
exist but with a sign opposite to that expected ( N$_{Bi_S}$= 3.97e
while N$_{Bi_L}$=3.55e). The quasi-absence of charge ordering can be
further appreciated on the projected band structure on Bi $s$ and O
$p$ levels plotted in Figure~\ref{fig1}.d. Between -12 to -6 eV, we
clearly observe four bands that have a dominant Bi $s$ character
rather indicating that both Bi$_S$ and Bi$_L$ cations have filled $6s$
levels. This is confirmed by our Wannier Functions (WFs) analysis
based on these 4 Kohn-Sham states that yields one $s$-like WF per Bi
cation (Figure~\ref{fig1}.c). In constrast, we observe two bands above
the Fermi level that have a dominant O $p$ character. Building again
the WFs for these two bands, we obtain two $s$-like WFs centered on
Bi$_S$ cations with large contributions coming from neighboring O
anions (Figure~\ref{fig1}.b). These two WFs are in fact the two oxygen
holes that are formed during the charge transfer from O to Bi$_S$
cations. This result is in close agreement with the work of Foyevtsova
{\em et al} and the electronic structure of SrBiO$_3$ can be
interpreted in terms of Bi$_L$ $6s^2$ and Bi$_S$
$6s^2\underline{L}^2$~\cite{BBO-Sawatzky} where the notation
$\underline{L}$ labels a ligand hole.

\begin{figure} 
\begin{center}
\resizebox{7.0cm}{!}{\includegraphics{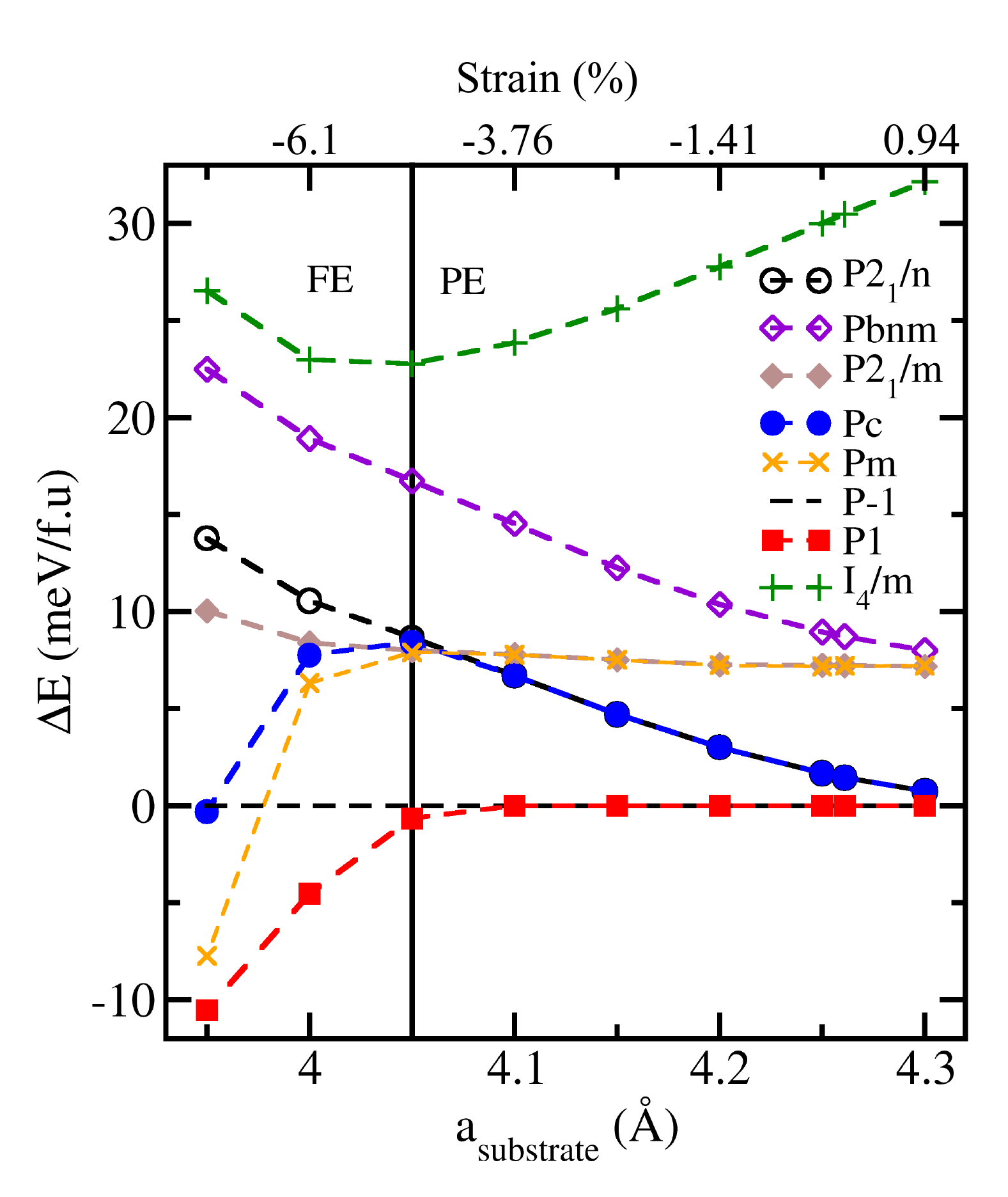}}
\end{center}
\caption{Energy differences (in meV/f.u) between the different
  metastable phases as a function of the substrate lattice parameter
  (in \AA, lower scale) or strain value (in \%, upper scale). The
  reference energy is set to the $P-1$ structure energy (dashed
  horizontal black line). The vertical line represents the boundary
  between ferroelectric (FE) and paraelectric (PE) phases. }
\label{fig2}
\end{figure}

Having now established that the structural and electronic properties
of SrBiO$_3$ are in good agreement with experiments and recent
theoretical works, we now turn our attention to spin-orbit related
phenomena. When including the spin-orbit interaction in the
calculation, we do not observe any sizable effect on the ground state
properties, rather suggesting a marginal SOI in this
material. However, the fact that SrBiO$_3$ is paraelectric in its bulk
hinders possible Rashba phenomena. Surprinsingly, we observe the
existence of a ferroelectric phonon mode at relatively low frequency
($\omega_{\text{FE}}$ $\simeq$ 50 cm$^{-1}$) in the bulk. Therefore,
through the coupling of this phonon mode with epitaxial
strain~\cite{RevueJV}, one may expect to unlock ferroelectricity in
SrBiO$_3$. This is reinforced by the ability of Bi-based oxide
perovskites to exhibit ferroelectricity in bulk or under epitaxial
strain. Since the pseudocubic lattice parameter of SrBiO$_3$ is rather
large (a$_{\text{pc}} \simeq$ 4.26~\AA), we only explored the effect
of a compressive strain in our simulations. Additionally, the SOI
having marginal effect on the structural aspect of the bismuthate, we
ignored it during geometry relaxation under strain.

\begin{figure}[h!]
\begin{center}
\resizebox{8.0cm}{!}{\includegraphics{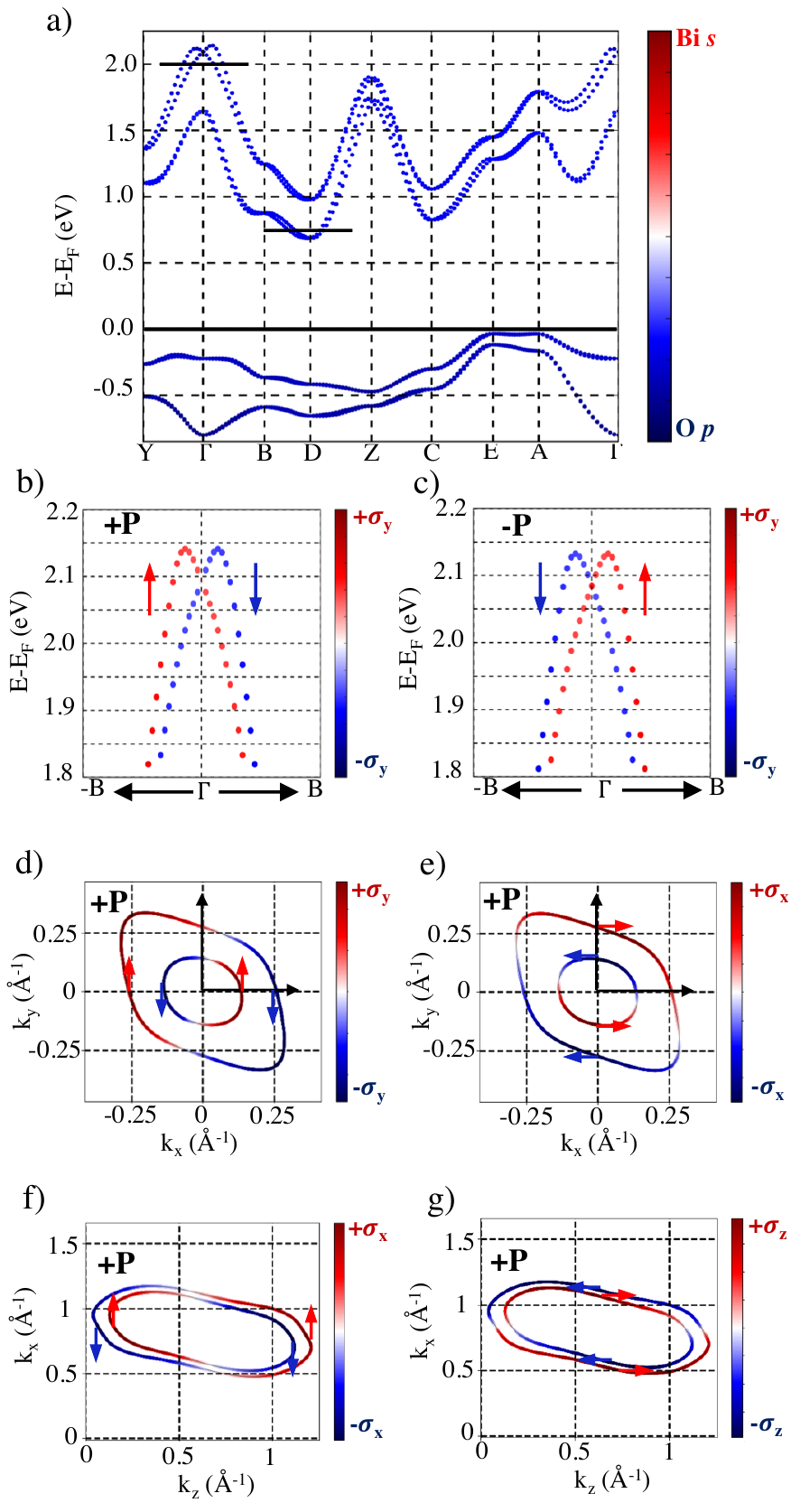}}
\end{center}
\caption{Electronic properties of the ferroelectric ground state of
  SrBiO$_3$ at 6.1 \% of compressive strain. a) Orbital projected --
  on O $p$ (blue character) and Bi $s$ (red character) states -- band
  structure in a monoclinic $P2_1/n$ setting. The Fermi level is set
  to 0 eV. b and c) Bands projected on the $\pm \sigma_y$ component
  along the $(-1/2, 0 , 0)$ to $(+1/2,0,0)$ direction for +P (c) or -P
  (d) spontaneous polarization. d and e) Spin textures in the
  $(k_x,k_y)$ plan obtained at E=+2 eV when projecting on the $\pm
  \sigma_{y}$ (d) or $\pm \sigma_{x}$ (e) spin component for a
  positive spontaneous polarization. f) and g) Spin textures in the
  $(k_z,k_x)$ plan obtained at E=+0.75 eV when projecting on the $\pm
  \sigma_{x}$ (f) or $\pm \sigma_{z}$ (g) spin component for a
  positive spontaneous polarization. }
\label{fig3}
\end{figure}

Figure~\ref{fig2} reports the energy of the most stable structures
found in the calculations as a function of the substrate lattice
parameter (lower scale)/compressive strain value (upper
scale)~\cite{noteStructure}. Between 0 and 5 \% of compression,
SrBiO$_3$ remains a paraelectric insulator adopting a $P\bar{1}$
structure described by a $a^-a^-c^+$ oxygen cage rotation patterns and
a breathing of O$_6$ groups. This structure is equivalent to the bulk
except that the in-phase oxygen cage rotation axis is aligned along
the substrate -- such a growth is sometimes referred to $c-P2_1/n$,
the notation $c-$ indicating that the $c$ axis is aligned along the
substrate~\cite{noteP-1}. Beyond 5\% of compression, the ground state
becomes ferroelectric and is associated with a $P1$ structure
exhibiting the aforementioned structural distortions plus a
spontaneous polarization in all three directions. Focusing on the
structure obtained at 6.1 \% of compressive strain -- {\em i.e.}
a$_{\text{substrate}}$=4~\AA --, SrBiO$_3$ develops a polarization of
12.65 $\mu$C.cm$^{-2}$ (P$_x$=9.64, P$_y$=2.73, P$_z$=-7.73 in a
pseudo $P2_1/n$ 20-atom unit cell setting, $c$ being the long axis)
approaching the value of the popular ferroelectric BaTiO$_3$.

When including the SOI in the ferroelectric phase, we now observe
several splittings of the ``oxygen hole bands'' located above the
Fermi level $E_F$ (see Figure~\ref{fig3}.a) in all reciprocal
directions due to the presence of polarization along the three
cartesian axes. The largest energy splitting is observed at the
$\Gamma$ point 2 eV above $E_F$, a smaller one being present at the
bottom of the conduction band at the D point 0.70 eV above
$E_F$. Turning our attention to bands around $\Gamma$, proper
inspection of their spin flavors along a (-B)-($\Gamma$)-(B) path
({\em i.e.} $-k_x$ to $+k_x$) reveals that their degeneracy is lifted
according to the spin direction as illustrated by
Figure~\ref{fig3}.b. Looking at spin textures in the ($k_x$,$k_y$)
plane represented in Figures~\ref{fig3}.d and~\ref{fig3}.e, we clearly
observe two energy contours with clockwise (outer contour) and
counter-clockwise (inner contour) spin-locking of electrons with
respect to their momentum following equation~\ref{eq1}. Consequently,
the ferroelectric phase of SrBiO$_3$ reached under compressive
epitaxial strain undergoes a Rashba interaction. Approximating the
Rashba coefficient by $\alpha_R = 2E_R/k_R$, where $E_R$ is the energy
splitting of the bands with respect to the high-symmetry position and
$k_R$ is the momentum offset, we estimate an unprecedented Rashba
coefficient of 8.397 eV.\AA~($E_R$=48.07 meV,
$k_R$=0.01145~\AA$^{-1}$) for the bands corresponding to the strongly
hybridized electronic structure between O $p$ and Bi $s$ states. At
the conduction band minimum, we do estimate rather large Rashba
coefficients of 0.944 eV.\AA~ and 0.474 eV.\AA~ along the D-Z and D-B
directions respectively (see Figures~\ref{fig3}.f and g for the spin
textures). Albeit smaller than values reported at the $\Gamma$ point,
these Rashba coefficients remain comparable to values observed in
BiAlO$_3$~\cite{BiAlO3} or in heavy metals such as (111) oriented
Bi~\cite{Bi111}.
\begin{figure} [h!]
\begin{center}
\resizebox{9.0cm}{!}{\includegraphics{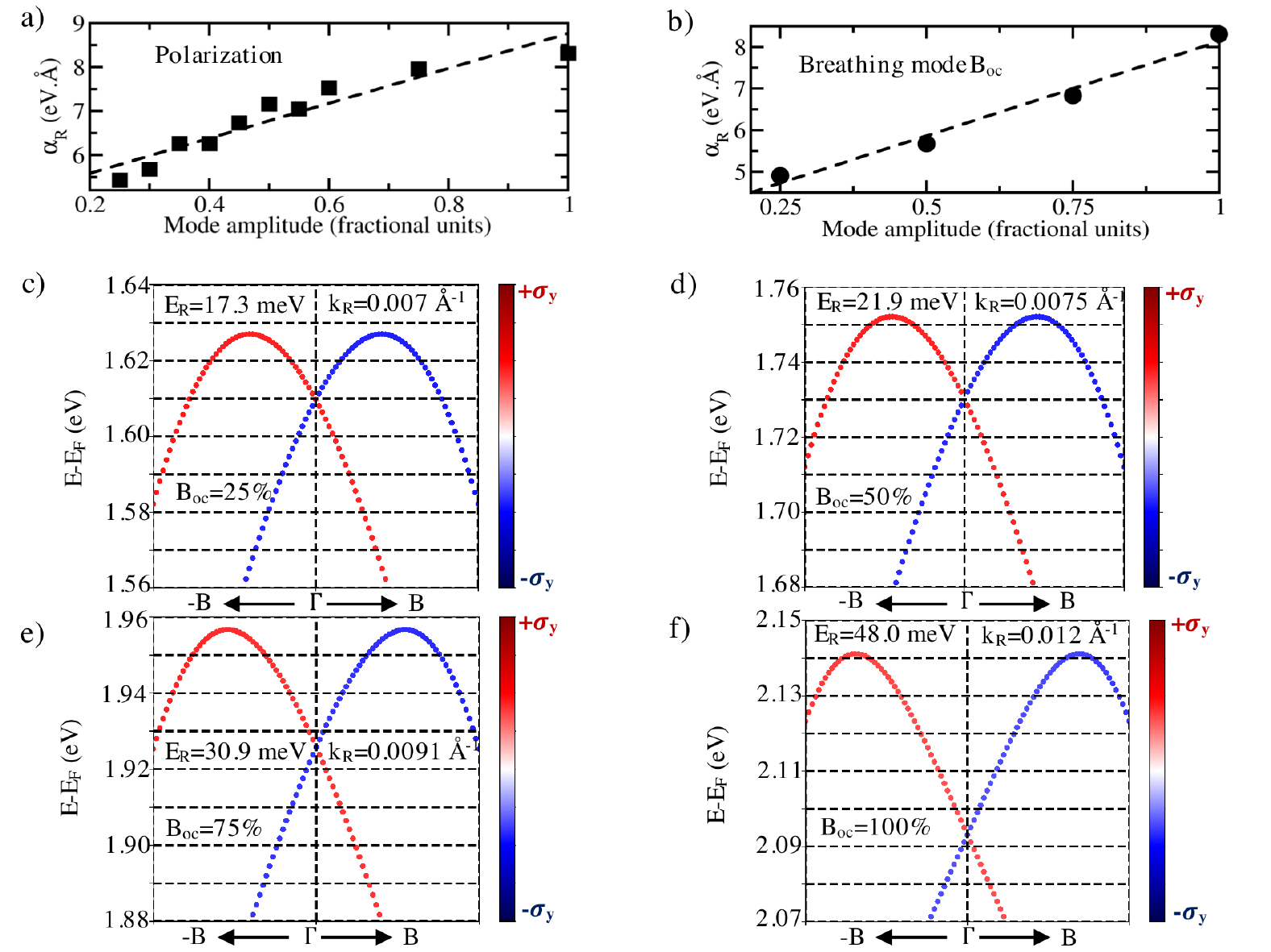}}
\end{center}
\caption{Evolution of the Rashba coefficient when freezing different
  amplitudes of the polar distortion (a) and the breathing of the
  oxygen cage octahedra (b).  c, d, e and f) Bands projected on the
  $\pm \sigma_y$ component along the (-1/2,0,0) to (+1/2,0,0)
  direction for different amplitudes of the breathing distortion. We
  emphasize that all other lattice distortions are fixed to their
  amplitude appearing in the ferroelectric ground state.}
\label{fig4}
\end{figure}

We now turn our attention to possible functionalization of the Rashba
interaction in this ferroelectric material. Firstly, as inferred by
Figures~\ref{fig3}.b and~\ref{fig3}.c, switching the polarization
reverses the spin textures therefore allowing their non-volatile
electrical control. By freezing different amplitudes of the
ferroelectric distortion while keeping other distortions to their
ground state amplitude, we observe a linear dependence of $\alpha_R$
with the ferroelectric mode amplitude (see Figure~\ref{fig4}.a), in
agreement with its dependence with the electric field
amplitude. Eventually, such functionalities were already proposed
theoretically in GeTe~\cite{GeTe} or in BiAlO$_3$~\cite{BiAlO3}. A key
aspect of the bismuthate is the strong relationship between the
breathing mode B$_{\text{oc}}$ and electronic properties. Although
B$_{\text{oc}}$ is found to open the band gap in the ferroelectric
phase upon its condensation in our simulations -- all other lattice
distortions are fixed to their ground state amplitude --, it also
controls the amplitude of the Rashba coefficient as inferred by
Figures~\ref{fig4}.b, c, d, e and f. Consequently, at odds with usual
assumptions, the Rashba coefficient is not only sensitive to the
spin-orbit strength $\lambda_{\rm SO}$ and the electric field
amplitude, but it can also be strongly related to structural
distortions altering the band dispersions and the level of
hybridization between the B site cation and O
states~\cite{LevelHybridization}. The latter observation is in close
agreement with the work of Bahramy {\em et al} highlighting that
hybridized electronic states around $E_F$ were a key aspect for large
$\alpha_R$~\cite{BiTeI-PRB84}.

Although the ferroelectric phase is reached at relatively large
compressive strain value, it might be achieved experimentally using
scandate substrates for instance. Additionally, the control of the
breathing mode may be unlocked by several external stimuli such as
strain engineering, octahedra connectivity~\cite{Rijnders} or with
hybrid improper ferroelectricity~\cite{RevueJV}.

In conclusion, we have predicted in this Letter unprecedented Rashba
spin splittings of oxygen hole bands in the covalent oxide perovskite
SrBiO$_3$. We have also demonstrated that the Rashba coefficient
$\alpha_R$ could be strongly related to lattice distortions
irrespective of the amplitude of the electric field or of the
spin-orbit interaction. The Rashba physics being at the core of the
most recent spin-orbitronics based applications, our work unveils new
pathways to optimize and control $\alpha_R$ through the subtle
interplay between lattice and electronic degrees of freedom appearing
in oxide perovskites.

\acknowledgments This work has been supported by the European Research
Council (ERC) Consolidator grant MINT under the contract
\#615759. Calculations took advantages of the Occigen machines through
the DARI project EPOC \#A0020910084 and of the DECI resource FIONN in
Ireland at ICHEC through the PRACE project FiPSCO. J. Varignon
acknowledges technical help from Adam Ralph from ICHEC.


\begin{thebibliography}{70}
\bibitem{noteStructure} We have tested several stuctures combining
  different oxygen cage rotations pattern and/or breathing distortion
  B$_{\text{oc}}$ and/or polar axes. Two growth orientation were used
  with either the $a$ and $b$ lattice parameters aligned at
  45$^{\circ}$ of the $a_{\text{sub}}$ and $b_{\text{sub}}$ substrate
  lattice parameters or the $c$ lattice parameter along the
  substrate. The eight most stable structures identified in the
  calculations are the following : $P2_1/n$ ($a^-a^-c^+$ +
  $B_{\text{oc}}$), $Pc$ ($a^-a^-c^+$ + $B_{\text{oc}}$ +
  $P_xP_yP_z$), $Pbnm$ ($a^-a^-c^+$), $P2_1/m$ ($a^+a^-c^-$), $Pm$
  ($a^+a^-c^-$ +$P_z$), $P\bar{1}$ ($a^+a^-c^-$ + $B_{\text{oc}}$) and
  $P1$ ($a^+a^-c^-$ + $B_{\text{oc}}$ + $P_xP_yP_z$).
\bibitem{noteP-1} The $P\bar{1}$ and $P1$ structures grow with the
  in-phase rotation axis along the substrate. We therefore have to use
  a 40 atoms unit cell to model the constraint in our
  calculations. Nevertheless, results regarding band structures and
  polarizations are reported with respect to a 20 atoms unit cell
  where the $c$ axis is the long axis of the perovskite.
\end{thebibliography}
\end{document}